\documentclass[acmtog]{acmart}
\AtBeginDocument{%
  \providecommand\BibTeX{{%
    \normalfont B\kern-0.5em{\scshape i\kern-0.25em b}\kern-0.8em\TeX}}}

\copyrightyear{2024}
\acmYear{2024}
\setcopyright{rightsretained}
\acmConference[TEI '24]{Eighteenth International Conference on Tangible, Embedded, and Embodied Interaction}{February 11--14, 2024}{Cork, Ireland}
\acmBooktitle{Eighteenth International Conference on Tangible, Embedded, and Embodied Interaction (TEI '24), February 11--14, 2024, Cork, Ireland}\acmDOI{10.1145/3623509.3635319}
\acmISBN{979-8-4007-0402-4/24/02}

\begin{document}

\title{Visions Of Destruction: Exploring Human Impact on Nature by Navigating the Latent Space of a Diffusion Model via Gaze}

\author{Mar Canet Sola }
\authornotemark[1]
\email{mar.canet@tlu.ee}
\orcid{0000-0001-5986-3239}
\affiliation{%
  \institution{Baltic Film, Media and Arts School, Tallinn University }
  \country{Estonia}
}

\author{Varvara Guljajeva }
\authornote{Both authors contributed equally}
\email{varvarag@ust.hk}
\orcid{0000-0002-0261-3121}
\affiliation{%
  \institution{Computational Media and Arts, Hong Kong University of Science and Technology (GZ)}
  \country{China}
}

\renewcommand{\shortauthors}{Mar Canet Sola \& Varvara Guljajeva}

\begin{teaserfigure}
  \includegraphics[width=\textwidth]{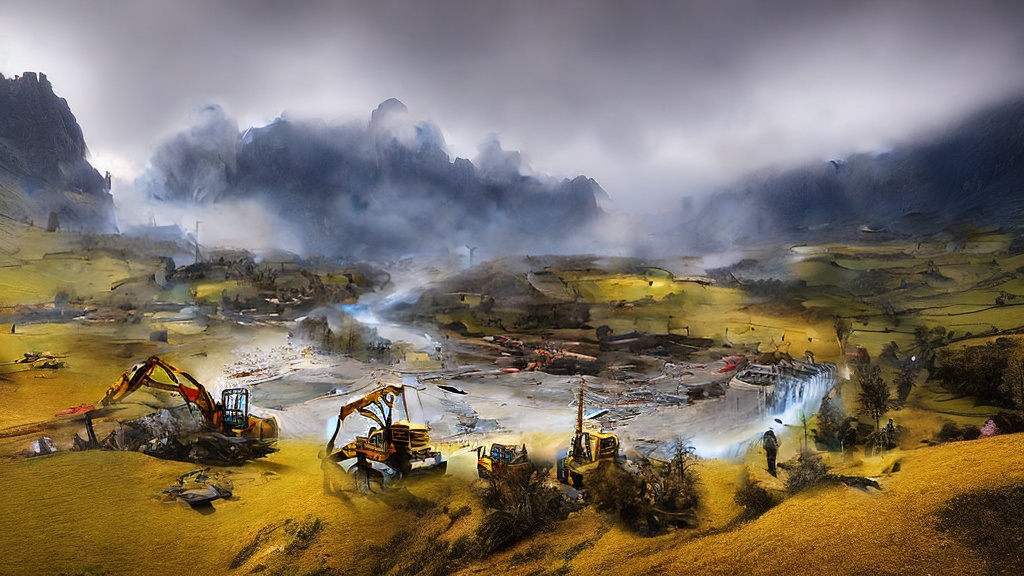}
     \caption{"Visions of Destruction" by Varvara \& Mar. A visual output when interacting with the installation.}
  \Description{}
  \label{fig:teaser}
\end{teaserfigure}

\begin{abstract}
This paper discusses the artwork "Visions of Destruction", with a primary conceptual focus on the Anthropocene, which is communicated through audience interaction and generative AI as artistic research methods. Gaze-based interaction transitions the audience from mere observers to agents of landscape transformation, fostering a profound, on-the-edge engagement with pressing issues such as climate change and planetary destruction.
The paper looks into early references of interactive art history that deploy eye-tracking as a method for audience interaction, and presents recent AI-aided artworks that demonstrate interactive latent space navigation.

\end{abstract}

\ccsdesc[500]{Applied computing~Media arts}
\ccsdesc[500]{Computing methodologies~Artificial intelligence}

\keywords{generative AI, interactive art, eye-tracking, Stable Diffusion, creative AI, natural user interface, gaze, Anthropocene}

\maketitle

\begin{figure*}[h]
  \centering
  \includegraphics[width=\linewidth]{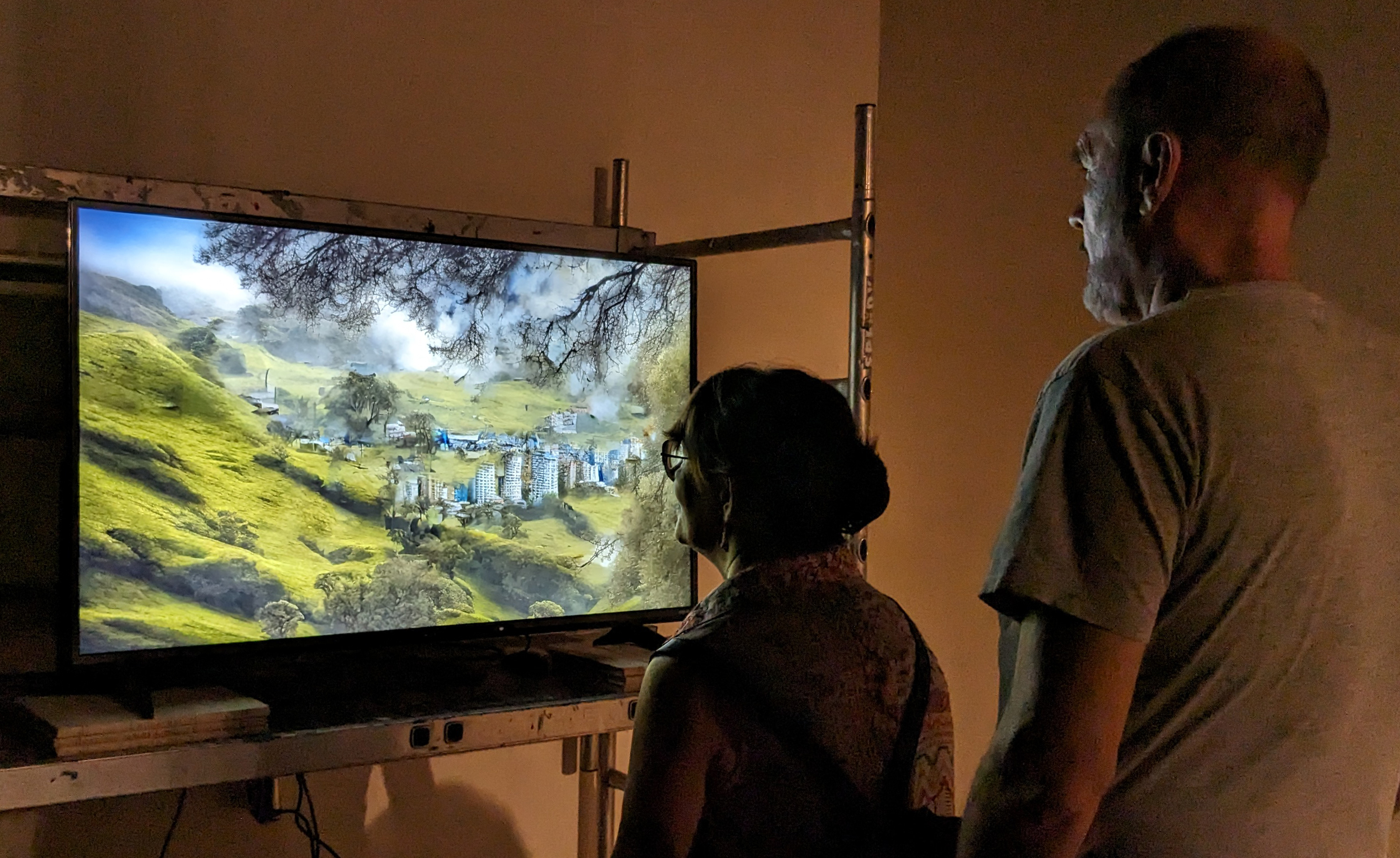}
    \caption{Audience engaging with the artwork "Visions of Destruction" (2023) by Varvara \&  Mar. }
  \Description{A participant engaging with the artwork 'Visions of Destruction' (2023) by Varvara \&  Mar.}
\end{figure*}

\section{Introduction}

The use of eye-tracking technology in art dates back to 1992 when the artwork "De-Viewer" by Art+com was created \cite{sauter2008interfaces} and shown in Ars Electronica. The work was done in response to the late 80s view of computers as tools rather than mediums; the artists transitioned from a brush to a mouse, using it similarly and using gaze to engage in a more mutual dialogue \cite{deviewer1992}. Three decades later, the same interaction methods can be realized with a discreet device attached to the screen's bottom, yet it remains a niche interface despite its high precision.

"De-viewer" aimed to ensure that by employing gaze as an interactive technique, the image would constantly change and never be seen the same way twice \cite{deviewer1992}. This concept is linked to the proposed artwork, combining interactive art and generative AI using Deep Learning (DL) based on Stable Diffusion. The idea behind "De-Viewer" significantly differed from "Visions Of Destruction". Joachim Sauter opposed mimicking the traditional art scene, especially concerning the media art community. Furthermore, audiences now engage with unexpected results of the generative AI's latent space rather than just a single image, as in 1992, proposing a new blaze of interaction aesthetics using generative AI models with an embodied interface. 

The early examples of interactive video installations using generative deep neural networks were projects used live camera input to create closed-circuit video systems, exemplified by Memo Akten's "Learning to See" (2017) \cite{akten2019learning} and by Mario Klingerman's "Uncanny Mirror" (2018) \cite{uncannymirror2018}. Nevertheless, a myriad of alternative interaction methodologies remain to be explored within the realm of interactive art, particularly in investigating the symbiosis with generative AI to engender artworks. In these prospective creations, participant interaction significantly influences the generative process, a frontier that remains relatively uncharted. A notable example is the art installation "Dream Painter"(2022) by artist-duo Varvara \& Mar \cite{canet2022dream, guljajeva2022dream, guljajeva2023artistic}, which using speech interaction method significantly contributes to this developing field by fostering a dynamic participant-artwork interaction where participant input matters in the output \cite{guljajeva2023explaining}. The burgeoning exploration of gaze interaction methodologies presented in this paper illuminates the potential for a more prosperous embodied method for dialogue between participants and generative AI within the sphere of interactive art. 

Building on Sauter's perspective \cite{deviewer1992}, it's essential to recognize creative forms from AI-interactive art as a distinct medium rather than just a tool. In a similar vein, Guljajeva emphasises the need for contextualization of AI-generated cultural content and discuss uprising neural avant-garde with its novel forms and processes \cite{guljajeva2021synthetic}. Hence, the aesthetic nuances born from the synergy of neural networks and human-machine interactions can yield new interpretations. These interpretations arise from the artistically designed system and the vast collective history embedded in datasets, such as LAION-5B. Generative AI accentuates generative art's fluidity and procedural attributes, underscoring its potential to encapsulate the dynamism inherent in cultural expressions derived from trained data. Within this medium, artists must craft systems and interfaces that navigate the latent space, ensuring meaningful connections to the artwork produced by having meaningful human control \cite{santoni2018meaningful} and interaction \cite{carpenter2019towards}.

\section{Concept}

\begin{figure*}[h]
  \centering
  \includegraphics[width=\linewidth]{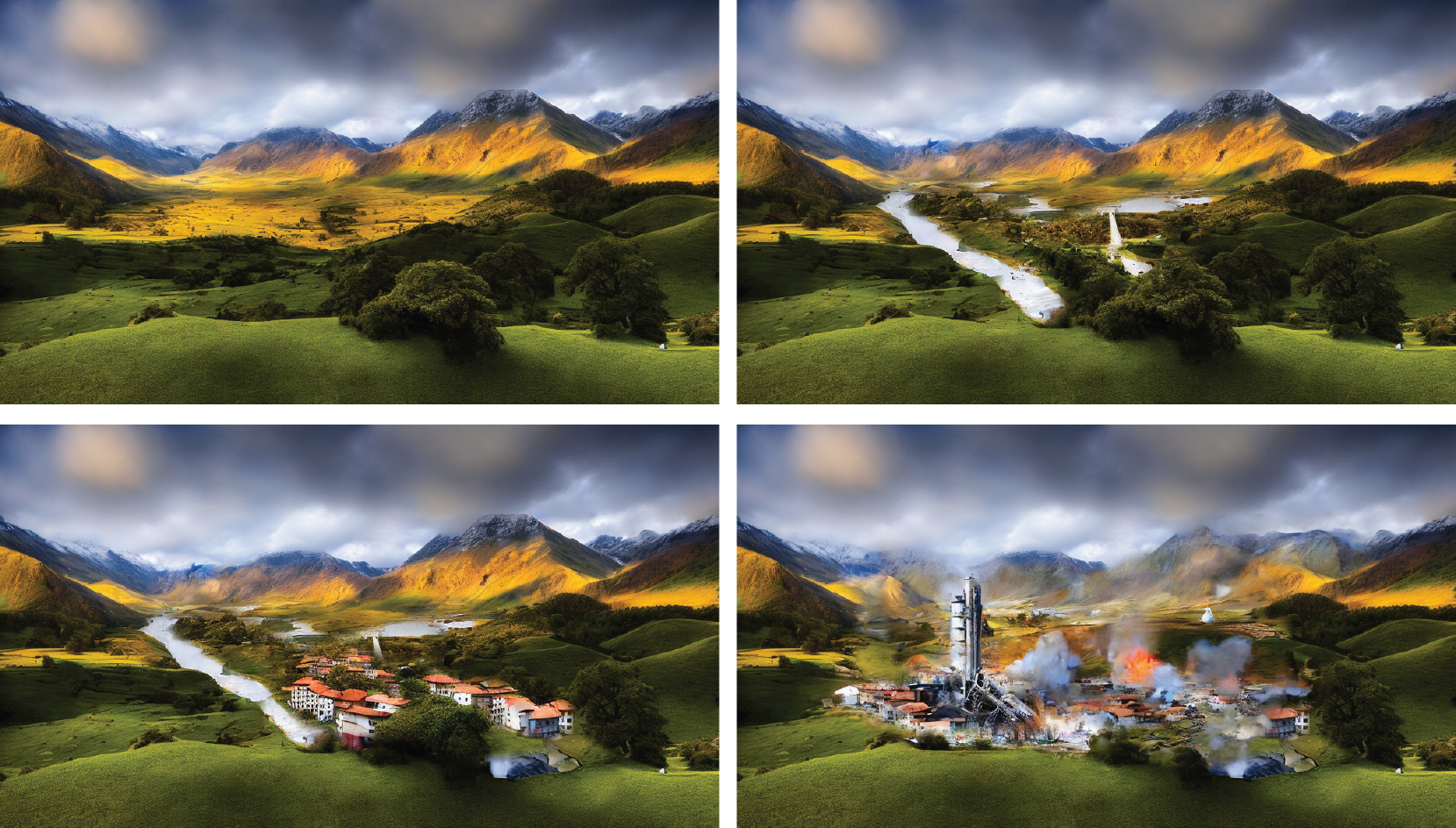}
    \caption{Sequence of images depicting the evolving landscape destruction as a participant interacts with the artwork. }
  \Description{Sequence depicting the evolving destruction of the landscape as a participant interacts with Varvara \& Mar's artwork 'Visions of Destruction' (2023).}
\end{figure*}

"Visions of Destruction" is a captivating interactive art piece that utilises eye-tracking technology to engage viewers with the pressing issue of environmental degradation (see Figure 1). A viewer's gaze, detected by an eye-tracking sensor, causes transformations in the landscape imagery (see Figure 2). Hence, merely by observing the digital scenery, the spectator induces dramatic changes at the points their gaze touches.
AI-generated 'beautiful landscapes', constructed by Stable Diffusion, present viewers with a romanticized version of nature derived from the collective human memory, as represented by a web-based training dataset. The piece operates in real-time, providing a unique experience for each viewer, symbolising the Anthropocene and the urgency to protect the natural environment. A viewer’s gaze is a metaphor for human presence and the irreversible actions leading to today's climate crisis.

Regarding audience interaction, an eye-tracker registers the gaze, which triggers an image change precisely where the audience’s eyes land. Using an array of pre-set prompts and inpainting with Stable Diffusion, viewers witness how pristine nature begins to deform before their eyes. Consequently, participants can reshape mountains, carve rivers, erect cities, and disrupt the initial idyllic nature, experiencing the metaphorical destruction and tension between technology and nature. When the eye-tracking detects no viewers, the landscapes begin a regeneration journey. Nature finds solace in this symbiotic dance between human presence and absence, its beauty flourishing. Additionally, the project brings interactivity into the realm of AI art.

The artwork effectively utilises generative models to emphasize the urgency of the climate crisis. By transitioning from serene landscapes to scenes of ecological devastation, it captures the stark realities of our evolving world. This aligns with the audience's crucial role in moulding our environment and highlights everyone’s duty to nature. As such, "Visions of Destruction" stands not only as artwork, but also as a call to action.

\section{Technical description}

"Visions of Destruction" is a screen-based interactive art installation that integrates an eye-tracking sensor for interactive audience engagement through a Natural User Interface (NUI) with a generative AI system. This immersive embodied experience blends generative AI with gaze-tracking, allowing viewers to intuitively navigate the latent space of the Stable Diffusion model in real-time and immerse in the artwork. Regarding technical realisation, the system captures a brief sequence of gaze data to form a mask area of the image, which serves as a changing area for the inpainting process of the Stable Diffusion model. Once the image is generated, the area in which the viewer's focused gaze on the landscape undergoes a smoothly animated transformation using FILM \cite{reda2022film} driven by one randomly selected prompt from a list of topic-relevant prompts predefined by the artist. We utilize generative AI algorithms to transform the areas of the landscape that the viewer has looked upon, simulating the gradual destruction of nature caused by human activity (see Figure 3). It metaphorically demonstrates how humans are responsible for the degradation of natural habitats by showcasing how the places we look upon gradually turn into polluted and destroyed landscapes. 

In essence, the viewer influences the trajectory of the AI image transformations rather than determining the exact visual content change. The design incorporates strategically crafted prompts that, when activated in random sequence, instigate alterations in the digital landscape at the exact spots where the audience gazes. This interactive mechanism ensures a unique and personalized artistic experience with each viewing. The system creates a contemporary interactive AI-driven visual collage using the gaze input data and affordances of the emergence of generative AI systems.

If the eye-tracking system does not detect viewers interacting with the artwork for a few seconds, it regenerates the landscape, transitioning to scenes of unspoiled, synthetic natural beauty. This transformation symbolizes nature's resilient power to regenerate after human impact. Remarkably, the generative AI consistently produces landscapes of stunning beauty, creating breathtaking vistas that, while purely imaginary, are surprisingly realistic.

The interactive and real-time nature of "Visions of Destruction" allows viewers to connect more intimately with the artwork's message and stimulates them to think more deeply about the environmental crisis. Eye-tracking technology enables us to create a personalized, unique and immersive experience that encourages viewers to contemplate the implications of their actions and take measures to protect the natural world. Overall, "Visions of Destruction" is a thought-provoking art piece that brings attention to the pressing issue of environmental degradation and raises awareness of the critical need for immediate action to protect our planet.

\section{Conclusion}

"Visions of Destruction" proposes a novel approach for engaging with a real-time generative AI system via an embodied interface facilitated by eye-tracking sensory technology. This method of tracking the viewer's gaze amplifies our artistic exploration of the Anthropocene and the impact of human activities on climate change. Moreover, this interactive element aids in demystifying AI technology, allowing audiences to experience the neural processes of the AI model directly. 

The artwork delves into the generative capabilities of AI systems when fused with an embodied interaction, creating different and meaningful experiences for each participant. Moreover, the gaze is an input for sensory technology, and simultaneously, an instrument to forward the artistic message to the public depicting humanity's footprint on the natural environment.

In the end, the artwork showcased here illustrates that we are witnessing a gradual decline in the hype surrounding DL technology by incorporating audience interaction techniques reminiscent of early interactive art. 

\section*{Biographies:}

Varvara \& Mar is an artist duo formed by Varvara Guljajeva and Mar Canet in 2009. Often duo’s work is inspired by the digital age. In their practice, they confront social changes and the impact of the technological era. In addition to that, Varvara \& Mar are fascinated by artificial intelligence, kinetics, audience participation, and digital fabrication, which are integral parts of their work. The artist duo has exhibited their art pieces in a number of international shows and festivals. Varvara \& Mar has exhibited at MAD in New York, FACT in Liverpool, Santa Monica in Barcelona, Barbican and V\&A Museum in London, Onassis Cultural Centre in Athens, Ars Electronica museum in Linz, ZKM in Karlsruhe, etc. And there research has been published in SIGGRAPH, SIGGRAPH Asia, IEEE VISAP, ACM TEI, EvoMUSART, ACM Creativity \& Cognition, ACM Multimedia, xCoAx, ISEA and more.

Dr Varvara Guljajeva is an Assistant Professor in Computational Media and Arts at the Hong Kong University of Science and Technology (Guangzhou). Previously, she held positions at the Estonian Academy of Arts and Elisava Design School in Barcelona. Her PhD thesis, “From Interaction to Post-Participation: The Disappearing Role of the Active Participant,” was selected as the highest-ranking abstracts by Leonardo Labs in 2020. 

Mar Canet Sola is a PhD candidate and research fellow at the CUDAN research group at BFM Tallinn University. He has a master’s degree from Interface Cultures at the University of Art and Design Linz and two degrees in art and design from ESDI in Barcelona and BSc (Hons) in computer game development from the University of Central Lancashire in the UK.

\begin{acks}
Mar Canet Sola and Varvara Guljajeva are the authors of the idea and conceptualization of 'Visions of Destruction'. Mar Canet Sola and Isaac Clarke did the software's technical development. Mar Canet Sola is supported as a CUDAN research fellow and ERA Chair for Cultural Data Analytics, funded through the European Union's Horizon 2020 research and innovation program (Grant No. 810961). 	
\end{acks}

\bibliographystyle{ACM-Reference-Format}
\bibliography{sample-base}

\appendix

\end{document}